\documentclass[aps,prl,twocolumn,showpacs]{revtex4}
\usepackage{amsmath,amsfonts,epsfig,graphicx,float}

\begin{document}

\title{Wrinkles Riding Waves in Soft Layered Materials}

\author{M. Abi Ghanem$^{1,2}$, X. Liang$^{2}$, B. Lydon$^{1}$, L. Potocsnak$^{1}$, T. Wehr$^{1}$, M. Ghanem$^{3}$, S. Hoang$^{1}$, S. Cai$^{2}$, and N. Boechler$^{1,2}$}

\affiliation{ 
$^1$ Department of Mechanical Engineering, University of Washington, Seattle, WA 98195 \\
$^2$ Department of Mechanical and Aerospace Engineering, University of California San Diego, La Jolla, CA 92093 \\
$^3$ Department of Aeronautics and Astronautics, University of Washington, Seattle, WA 98195
}

\begin{abstract}
The formation of periodic wrinkles in soft layered materials due to mechanical instabilities is prevalent in nature and has been proposed for use in multiple applications. However, such phenomena have been explored predominantly in quasi-static settings. In this work, we measure the dynamics of soft elastomeric blocks with stiff surface films subjected to high-speed impact, and observe wrinkles forming along with, and riding upon, waves propagating through the system. We analyze our measurements with large-deformation, nonlinear visco-hyperelastic Finite Element simulations coupled to an analytical wrinkling model. The comparison between the measured and simulated dynamics shows good agreement, and suggests that inertia and viscoelasticity play an important role. This work encourages future studies of the dynamics of surface instabilities in soft materials, including large-deformation, highly nonlinear morphologies, and may have applications to areas including impact mitigation, soft electronics, and the dynamics of soft sandwich composites.
\end{abstract}

\maketitle

Instability patterns are often observed in stiff films on soft elastic substrates subject to compressive stresses \cite{Gao2012}. These patterns include wrinkles \cite{Allen1969,Whitesides1998,Brau2010}, ridges \cite{Zhao2014,Hutchinson2015}, and folds \cite{Cerda2008,Stone2011,Kim2013}. The mechanics of such surface instabilities have been the subject of significant interest due to their common appearance in nature and their applications in areas such as scalable nanomanufacturing \cite{Zhao2016} and morphing structures \cite{Reis2014}. 

Among the range of observed elastic surface instabilities, wrinkling patterns have been most intensively studied. The profile of such wrinkles are typically described by a sinusoidal function with a wavelength that is mainly determined by the stiffness mismatch between the stiff film and soft substrate underneath, and the thickness of the film \cite{Whitesides1998}. While this setting has been widely explored in quasi-static regimes, wherein inertial effects are ignored, the dynamic formation and propagation of such surface wrinkles in soft, viscoelastic, layered elastomers subjected to high-speed impact has hitherto not been studied. Many works have explored the dynamics of related systems prone to elastic instabilities ranging from studies of dynamic buckling of thin bars \cite{Lindberg1965,Villermaux2005,Hutchinson2006} to Schallamach waves \cite{Schallamach}, and studies (see Ref. \cite{Villermaux2013} and references therein) including the dynamics of wrinkling instabilities of thin, freestanding elastic sheets \cite{Villermaux2009} and bands \cite{Villermaux2007}, floating membranes \cite{Duchemin2016}, and filaments in viscous fluids \cite{Kudrolli2017}. Recent works have also explored the dynamics of layered composites containing stiff inclusions in soft, viscoelastic matrices subjected to strain rates of up to $10^{-1}$ s$^{-1}$ \cite{Rudykh2016}, and axial dynamic pulse buckling in sandwich composite plates \cite{Simonetta1997, Ji2008}. In contexts directly related to wrinkling of stiff films on viscoleastomeric bases, recent studies have also theoretically explored the dynamics of wrinkle growth and coarsening \cite{Huang2002a,Huang2002b,Huang2005,Im2005,Im2006}, as well as experimentally studied the slow ($\sim300$ s) growth and reorganization of folds and wrinkles under biaxial compression \cite{Stone2011} and changing compression direction \cite{Shimomura}, respectively, slip dynamics of ripple dislocation \cite{Chou2011}, evaporation-driven wrinkle growth \cite{Rowat2017}, and nanoscale anisotropic wrinkle growth under the presence of ion bombardment \cite{Giordano2018}. 

\begin{figure*}[t]
\begin{center}
\includegraphics[width=15 cm]{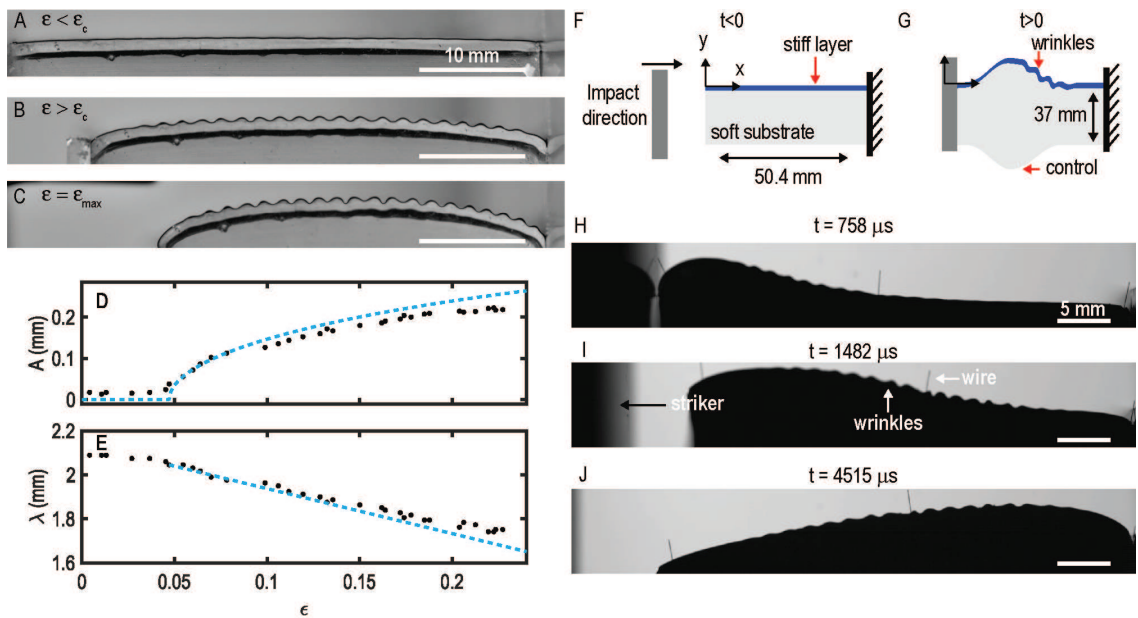}
\end{center}
\caption{\label{Figure1} [Color online] Quasi-static compression tests of the elastomeric block showing the side containing the stiff surface film for strains $\epsilon$ (A) less than and (B) greater than the critical wrinkling strain $\epsilon_c$, and (C) equal to the maximum strain tested $\epsilon_{max}$. Wrinkling amplitude (D) and wavelength (E) as a function of strain for the tests shown in (A-C). Black dots denote the measured data, and the blue dashed lines the quasi-static wrinkling model of Eq.~\ref{wavelengthandamp} using the fitted parameters. Schematic of the experiment, before (F) and after (G) impact. High-speed video images taken at times $t=758$ $\mu$s (H), $1482$ $\mu$s (I), and $4515$ $\mu$s (J) after contact between the striker and the sample for an impact velocity of $V_S \approx 10$ m/s.}
\end{figure*}

In this work, we study the dynamics of surface wrinkle formation and propagation in a soft elastomer block containing a stiff surface film due to high-speed plate impact (simulated strain rates over $500$ s$^{-1}$ and plate velocities approaching $\sim25\%$ of wave speeds observed in the block). The plate travels such that its velocity vector lies within the plane of the film, and the impact launches a large-deformation compression wave in the block that induces progressive wrinkle formation. We measure the evolution of the wrinkles using high-speed video and compare the measured dynamics with those of the opposite side of the same block (which does not contain a surface film). By tracking the out-of-plane motion of the surfaces and the two-dimensional motion of several surface particles in each case, we see that inertial (wave propagation) effects play a major role in the substrate and drive the formation of wrinkling instabilities. We observe that the wrinkle formation speed correlates with, but is slightly slower than, the speed of the compression wave in the block, the trajectory of the tracked surface particles are nearly the same for both cases (with and without the film), and the motion of individual wrinkles follow the trajectory of the tracked surface particles. In other words, we observe wrinkles to form with, and ``ride'' the waves propagating through the substrate. We then model the motion of the substrate using Finite Element (FE) simulations that incorporate large-deformation mechanics, visco-hyperelasticity, and inertial effects. The resulting surface wrinkle dynamics are modeled by applying a quasi-static wrinkling model that uses strain in the longitudinal direction, averaged over the thickness of the block, in the FE simulations as an input. In this approach, wrinkles form on the surface once a critical level of strain is induced and the wrinkle phase is tied to the material points of the substrate surface. Using this simplified model, we find good agreement between the measured and simulated wrinkling dynamics, with a few exceptions. In our model, the speed of wrinkle formation appears to match the wave speed in the simulated substrate, whereas in the experiments, the wrinkle formation speed is relatively slower than the measured wave speed in the substrate. We also find that while our wrinkling model provides a good match to the quasi-static compression tests of the sample, the wrinkle amplitudes in the dynamic compression tests are approximately half of those found in simulations. We suggest that these differences can be explained in the context of inertial and viscoelastic effects. 

\section{Results and Discussion}

Our elastomeric sample that is prone to surface instabilities consists of a polydimethylsiloxane (PDMS) block ($L_0=50.4$ mm in length, with a square cross-section of height $H=37$ mm) that contains a stiff PDMS surface film adhered to one side ($h=140$ $\mu$m thick, as measured by optical profilometry). Additional sample fabrication details can be found in the supplementary information. A photograph of the side of the sample containing the stiff film and the resulting, well-known, wrinkling pattern that occurs as the sample is quasi-statically compressed is shown in Figs.~\ref{Figure1}A-C. At the largest strain tested $\epsilon_{max}$, we note the beginning of period doubling in the wrinkle pattern \cite{Brau2010}. Within several millimeters of the boundaries, particularly near the left boundary at intermediate strains, the wrinkling pattern appears suppressed. We attribute this to confinement caused by the boundaries or to sample inhomogeneity. By applying an edge detection algorithm to images taken during the compression test, we obtain the surface profile as a function of the average compressive strain along the tangential direction of the surface $\epsilon$ = $1 - s(L)/L_0$, where $s$ is the arc length of the curved surface, $L$ is the compressed length of the sample, and $\epsilon$ is taken as positive in compression for convenience. Taking a Hilbert and Fourier transform of the surface profile, we obtain the amplitude $A$ and wavelength $\lambda$ of the surface wrinkles, respectively, and plot them as a function of applied strain $\epsilon$, as is shown in Figs. ~\ref{Figure1}D and ~\ref{Figure1}E. The measured dependence is well captured using the following nonlinear quasi-static wrinkling model \cite{Brau2010}:

\begin{figure*}[t]
\begin{center}
\includegraphics[width=17.9 cm]{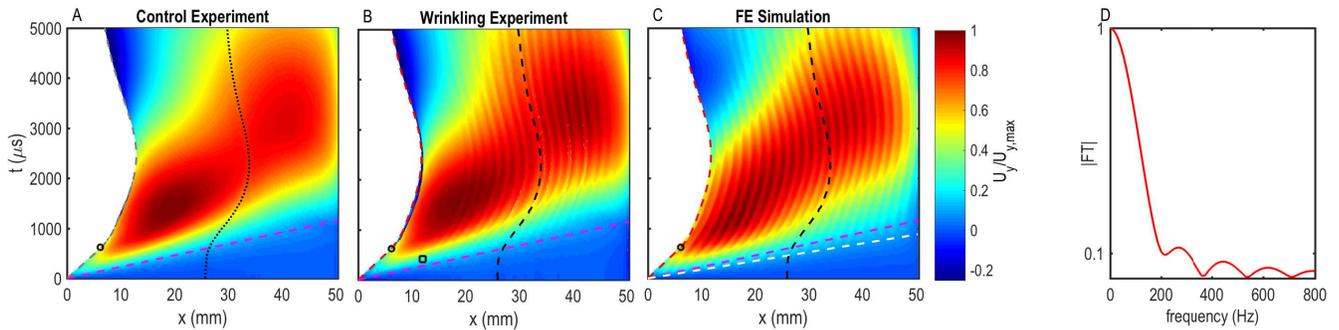}
\end{center}
\caption{\label{Figure2} [Color online] (A-C) Out-of-plane surface displacements $U_y$ (in a fixed ``laboratory'' reference frame), normalized by the maximum displacement in each sample $U_{y,max}$, in response to plate impacts of $V_S\approx10$ m/s velocity. Displacement measured for two experiments where high-speed images were taken on the side of the sample without (A) and with (B) a surface film (corresponding to the images in Fig.~\ref{Figure1}H-J). Displacement simulated (C) using a FE model of the substrate in combination with an analytical quasi-static wrinkling model that uses the simulated, depth-averaged, normal strain in the longitudinal direction $\epsilon_{xx}$ as an input. The dashed gray and red lines denote the identified longitudinal position of the sample boundary in the control and wrinkling experiments, respectively, the black circle the point of separation between the striker plate and the sample boundary, and the black square marker the approximate point of initial wrinkle formation. The dashed magenta and white lines respectively denote the identified wave speed in the substrate in the experiments $c$ and simulation $c_S$. The dotted and dashed black lines trace the lateral position of the middle probe wire in the control and wrinkling experiments, respectively. The scale bar is truncated at $U_y/U_{y,max}=-0.25$. (D) The Fourier transform spectrum of the longitudinal motion of the left boundary denoted by the dashed red line in panels B and C.}
\end{figure*}

\begin{align}
\begin{split}
\lambda & =  \lambda_c\left[1-(\epsilon-\epsilon_c)]\right., \\
A & =  \lambda_c \left[ \frac{\sqrt{\epsilon-\epsilon_c}}{\pi} \left(1-   \frac{3(\epsilon-\epsilon_c)}{8}   - \frac{17(\epsilon-\epsilon_c)^2}{128} \right )\right],\\
 \label{wavelengthandamp}
\end{split}
\end{align}

\noindent where $\lambda_c = \lambda(\epsilon_c)$ is the wrinkle wavelength at the critical strain $\epsilon_c$ at which the wrinkles begin to appear and $\lambda$ is undefined for $\epsilon<\epsilon_c$. We choose $\epsilon_c = 0.043$, which corresponds to the first point in Fig.~\ref{Figure1}D that exhibits a significant increase in amplitude. At the critical strain, we observe a critical wrinkling wavelength of $\lambda_c = 2.0$ mm. In Figs.~\ref{Figure1}D and ~\ref{Figure1}E, wrinkles with non-negligible amplitudes can be seen even before the identified critical strain. We note that the sample was subjected to multiple cycles of quasi-static and dynamic compressions prior to the experiments shown herein. As such, the results shown in this work should be considered representative examples. We suggest that the pre-existing wrinkles are the result of small, irreversible deformation induced during the prior quasi-static and dynamic testing of the sample. Such memory effects in systems exhibiting surface instabilities have been observed previously \cite{ShimomuraMemory}. The presence of pre-existing wrinkles likely initializes wrinkles in the same location for each test, however, we do not expect that it significantly affects the observed wrinkling amplitude or wavelength. 

High-speed impact experiments are performed using a spring-loaded drop tower that launches an aluminum plate at the sample, as is shown in Fig.~\ref{Figure1}F. The plate impacts the sample at a velocity of $V_S\approx10$ m/s and launches a large-deformation compression wave that induces progressive wrinkle formation on the layered side of the sample, as is illustrated in Fig.~\ref{Figure1}G. The plate is halted by metal stoppers after compressing the sample by $l\approx6.3$ mm, the equivalent of an applied static strain of $\epsilon_S=0.12$ (not accounting for surface curvature) \cite{Footnote1}. Assuming constant striker velocity until it hits the stoppers, we estimate strain rates exceeding $\dot{\epsilon_S}\approx\epsilon_S V_S/l=200$ s$^{-1}$. The resulting dynamics are measured using high-speed video (video files are included with the supplementary information). As shown in Fig.~\ref{Figure1}F, we define an Eulerian coordinate system where $x$ corresponds to longitudinal or in-plane motion (with respect to the surface film), and $y$ denotes lateral or out-of-plane motion. Figures ~\ref{Figure1}H-J show snapshots of the layered surface at several times after the impact. The edge shown in the dynamic tests of Fig.~\ref{Figure1}H-J is not necessarily the same as that shown in the quasi-static tests as shown in Fig.~\ref{Figure1}A-C, although it is the same face. Small probe wires that were used to follow the motion of multiple surface particles can also be seen. For the image at time $t=758$ $\mu$s, we observe wrinkling instabilities beginning to form at the front edge of the propagating compressive pulse, while the region in front of the pulse, where the wave has not yet arrived, remains mostly flat.  At $t=1482$ $\mu$s, new wrinkles progressively appear and the existing wrinkles grow in amplitude as the wave travels further into the sample. By $t=4515$ $\mu$s, the pulse has reached the opposite boundary. In Figs.~\ref{Figure1}H-J, we note the appearance of localized regions of aperiodicity in the surface wrinkling, which we attribute to a combination of disturbance caused by the presence of the probe wires and inhomogeneity in the strain field. Following this experiment, we perform a second experiment on the same sample, using about the same striker velocity \cite{Footnote1}, but record the dynamics of the opposite side of the sample, which does not contain a surface layer. We henceforth refer the prior experiment as the ``wrinkling'' case, and the latter experiment, without a surface film, as the ``control'' case.

We extract the out-of-plane surface displacement $U_y$ at each time captured in the high-speed video using an edge detection algorithm and plot the resulting spatiotemporal evolution (additional signal processing details for both the quasi-static and dynamic experiments can be found in the supplementary information). Figures ~\ref{Figure2}A and ~\ref{Figure2}B show the dynamics of $U_y$ measured in the laboratory reference frame for the control and wrinkling cases, respectively. In these and the subsequent spatiotemporal plots, any points detected to the left of the identified moving free boundary were discarded. The wavefront speed in the control case is well described by a constant speed of $c$ = 44 m/s, which we find by fitting a line (the dashed magenta line) to the time at each position where $U_y$ reaches $10\%$ of the maximum displacement. We observe a region of large curvature near the impacted edge at early times, which subsequently spreads throughout the sample. The in-plane trajectories of the middle probe wires for the control and wrinkling cases are shown by the dotted and dashed black lines, respectively. For the wrinkling case, we observe qualitatively similar surface displacement dynamics as the control case, but with the addition of small, superimposed wrinkles. The wrinkles begin to form at  $t\simeq400~\mu$s from the striker impact and $\sim12$ mm from the sample boundary, and their motion closely follows the surface particle motion identified by the probe wire. The wrinkles also appear to form slower than the wavefront speed $c$. The spatiotemporal response for additional dynamic impact tests, repeated with approximately the same parameters as in Fig. \ref{Figure2} but on different edges and orientations of the sample, are shown in Fig. S1. 

We model the dynamics of the elastomeric foundation via FE simulations using the commercial software Abaqus/Explicit \cite{Abaqus} and consider large-deformation, geometrically nonlinear effects. We consider a two-dimensional plane strain problem and model the foundation as a rectangular block made of visco-hyperelastic material of length $L_0$, height $H$, and density of $\rho = 965$ kg/m$^3$. The hyperelasticity is modeled as a Neo-Hookean material with shear modulus $\mu_0=44$ kPa and Poisson's ratio $\nu=0.49$, where the strain energy density function is $W=\frac{1}{2}\mu(\Lambda_1^2+\Lambda_2^2+\Lambda_3^2-3)+\frac{1}{2}K(J-1)$, $K$ is the bulk modulus, $\Lambda_i$ are the principle stretches, and $J=\Lambda_1 \Lambda_2 \Lambda_3$ represents the volumetric change of the material. The viscoelasticity is modeled using a Standard Linear Solid (SLS) element material \cite{LakesBook}, with a dynamic shear modulus that relaxes such that $\mu(t)$ = $\mu_0$ + $\mu_1e^{-t/\tau}$, where $\mu_1$ = 3.35 MPa, $\tau = 90$ $\mu$s, and subscripts $0$ and $1$ denote the long- and short-term shear moduli, respectively. The block is dynamically compressed by specifying the horizontal and vertical displacements $U_x$ and $U_y$ measured as a function of time for the left edge of the sample in the wrinkling case. The horizontal displacement of the boundary is the same for all nodes along the left edge, and the vertical displacement is linearly interpolated, from a maximum at the top and bottom left corners to zero at the axis of symmetry of the block. The nodes along the right boundary (at $x$ = 50.4 mm) are horizontally and vertically fixed ($U_x$ = $U_y$ = 0). Additional FE simulation details can be found in the supplementary information. 

We simulate the wrinkle dynamics using a quasi-static wrinkling model based on Eq.~\ref{wavelengthandamp}, where the depth-averaged normal strain in the longitudinal direction $\epsilon_{xx}$ is used as an input, such that the strain in Eq.~\ref{wavelengthandamp} is set to be $\epsilon = \epsilon_{xx}$. The strain $\epsilon_{xx}$ obtained from the simulations corresponds to the nominal strain (where $\epsilon_{xx}=1-\Lambda_x$, $\Lambda_x$ is the stretch in the $x$ direction, and compression is defined as positive $\epsilon_{xx}$ as before), is averaged over a prescribed depth $D=H/2$ (with coordinates defined in terms of the undeformed state), and varies with longitudinal position. Averaging over the depth is a large assumption, as the strain distribution under such dynamic loading is highly heterogeneous (see Fig. S2). The wrinkle profile is defined as a sinusoid in the undeformed material reference frame with a constant wavelength of $\lambda_c$ \cite{Villermaux2009}, a phase such that the amplitude is maximum at the left boundary, and an amplitude that depends on the depth-averaged strain $\epsilon_{xx}$, such that the wrinkles begin to form when $\epsilon_{xx}=\epsilon_c$. Since the wrinkle motion follows the same trajectory as the tracked surface particles in both the control and wrinkle cases (shown by the wire profiles in Figs. ~\ref{Figure2}), we assume that the deformation following the impact changes the wrinkling amplitude, but not the phase. The wrinkle phase is tied to its initial pre-impact material point, such that the wrinkles move with the deformed substrate and areas of high compression result in an apparent shift in the wrinkling wavelength in the laboratory reference frame. Using $x'$ and $y'$ to denote the horizontal and vertical position in the undeformed material reference frame, respectively, the wrinkled surface profile is described as:

\begin{align}
\begin{split}
x(x',t)  = x' + &U_x(x',t) \\ - &\left( A(\epsilon(x',t))\cos \left(\frac{2\pi x'}{\lambda_c}\right) \right) \sin \alpha(x',t),\\
y(x',t)  = y'+&U_y(x',t) \\ + &\left( A(\epsilon(x',t))\cos \left(\frac{2\pi x'}{\lambda_c}\right) \right) \cos \alpha(x',t),\\
 \label{Eq2}
\end{split}
\end{align}
where $A$ is the wrinkling amplitude described by Eq. ~\ref{wavelengthandamp} and $\alpha$ is the angle of surface tangent from the x-axis. 

Figure ~\ref{Figure2}C shows the simulated spatiotemporal evolution of the out-of-plane surface displacement of the block in the laboratory reference frame, accounting for the presence of wrinkles using Eq.~\ref{Eq2}. Ignoring the high-frequency wrinkles, the simulated low-frequency, out-of-plane displacement dynamics appear to agree reasonably well with those in the experiments. One noticeable difference is the simulation has a faster wavefront speed of $c_S=57$ m/s (identified using the same criteria as in Fig.~\ref{Figure2}A), as is denoted by the dashed white line in Fig.~\ref{Figure2}C, as well as a faster wrinkle formation speed. Both speeds $c$ and $c_S$ are within the range expected given the simulated material properties. For instance, considering the substrate as a one-dimensional (1D) viscoelastic bar, we expect sound speeds ranging from $c_0=\sqrt{E_0/\rho}=11.7$ m/s to $c_1=\sqrt{(E_0+E_1)/\rho}=102$ m/s, where the elastic modulus $E_m$ is related to the shear modulus such that $E_m=2\mu_m(1+\nu)$, the subscript $m$ denotes the long or short-term moduli, and any viscoelastic variation of $\nu$ is ignored. The upper bound of $c_1$ is likely overestimated, due to effects such as the interplay of lateral inertia with the frequency-dependent viscoelastic response of the substrate (as has been explored in Ref. \cite{Harrigan2010}). Using Love's equation for waves in an elastic bar accounting for lateral inertia effects \cite{Love}, a maximum frequency supported by the bar at short wavelengths can be estimated as $f_n=c_n \sqrt{6}/(2 \pi \nu H)$, where $n=\{0,1\}$ and $f_0=250$ Hz and $f_1=2.20$ kHz. Consistent with this range, a Fourier transform of the left boundary longitudinal displacement history (shown in Fig.~\ref{Figure2}D) suggests that the frequencies injected into our system are less than $\sim 400$ Hz. Assuming a sinusoidal applied strain, the dynamic modulus of an SLS material can be represented as $|E^*|=|E'+iE''|$ and the loss tangent as $\tan(\delta)=E''/E'$, where the storage modulus is $E'=E_0+E_1\omega^2\tau^2/(1+\omega^2\tau^2)$, the loss modulus is $E''=E_1\omega\tau/(1+\omega^2\tau^2)$ \cite{LakesBook}, and $\omega=2\pi f$ is the angular frequency. Using the identified maximum frequency from Fig.~\ref{Figure2}D, we estimate a maximum 1D wave speed of $c^*=\sqrt{(|E^*|/\rho)}\sec(\delta/2)=60$ m/s.

\begin{figure}[t]
\begin{center}
\includegraphics[width=9 cm]{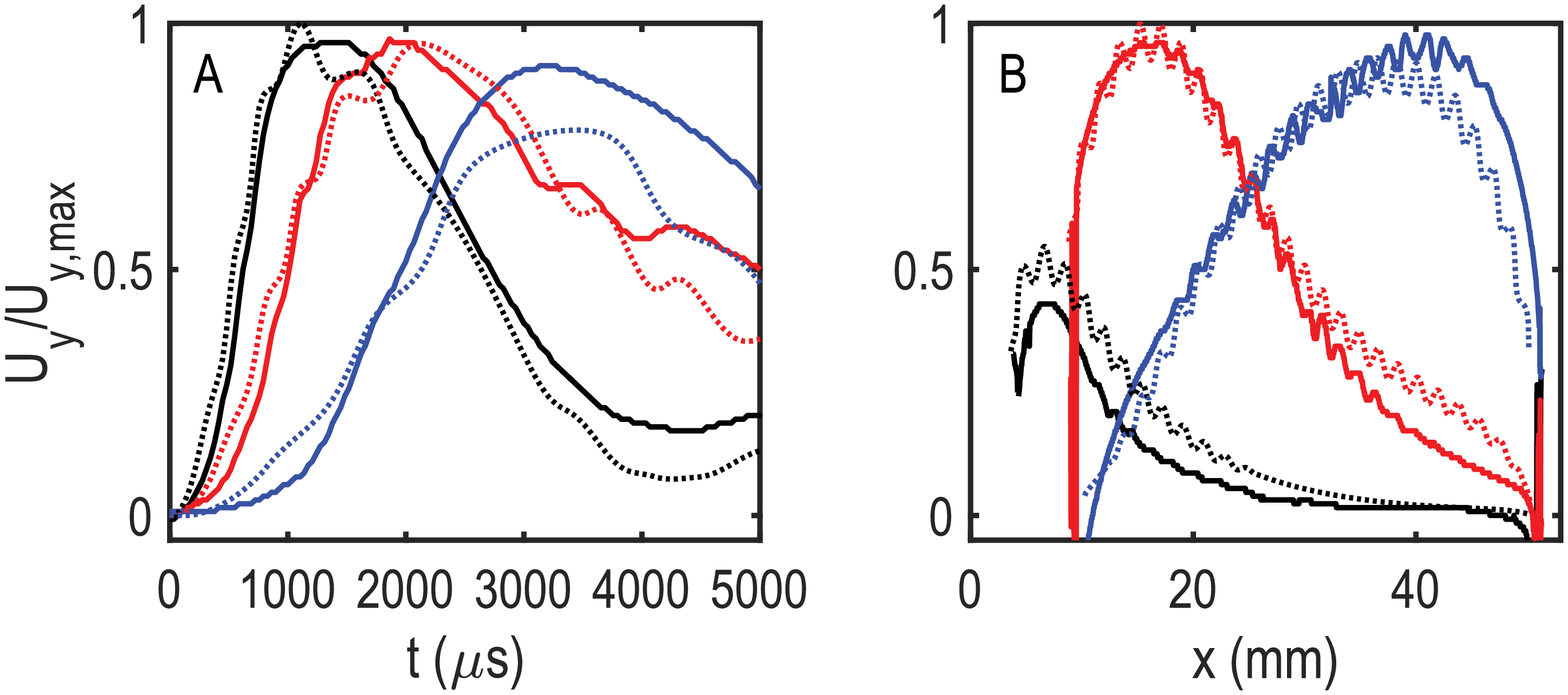} 
\end{center}
\caption{\label{Figure3} [Color online] Measured (solid lines) and simulated (dotted lines) normalized out-of-plane surface displacement for several positions and times selected from the data in Fig.~\ref{Figure2}B and ~\ref{Figure2}C. (A) Displacements as a function of time, for positions $15$ mm, $25$ mm, and $45$ mm from the left boundary, denoted by the black, red, and blue lines, respectively. (B) Displacement profiles at times $0.38$ ms, $1.2$ ms, and $3.4$ ms after striker contact, also denoted by the black, red, and blue lines, respectively. The spike in the red curve in panel B at $50.4$ mm is due to detection of the sample boundary.}
\end{figure}

\begin{figure*}[t]
\begin{center}
\includegraphics[width=17.9 cm]{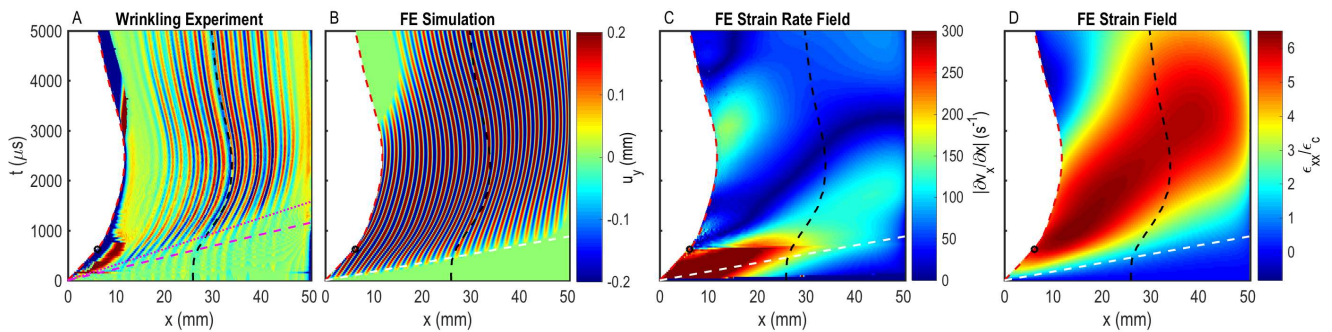}
\end{center}
\caption{\label{Figure4} [Color online] Spatiotemporal evolution of the (A) measured and (B) simulated out-of-plane surface wrinkling (in a fixed ``laboratory'' reference frame) after removal of the low-frequency compressive pulse traveling in the substrate (denoted as $u_y$). The dotted magenta line is the experimentally identified speed of wrinkle formation and the dashed magenta line the experimentally identified sound speed for the control experiment case. (C) Magnitude of the strain rate $|\dot{\epsilon_{xx}}|=|\partial v_x/\partial x|$ obtained from the FE simulations, where $v_x$ is the particle velocity in the $x$ direction. The strain rate is truncated to a maximum of $300$ s$^{-1}$. (D) Normalized spatiotemporal evolution of the strain $\epsilon_{xx}/\epsilon_{c}$ obtained from the FE simulations. The dashed white line in panels B-D is both the sound speed and speed of wrinkle formation identified in the simulations. Across all panels, the dotted and dashed black lines, the dashed red lines, and the black circle denote the same quantities as in Fig.~\ref{Figure2}.} 
\end{figure*} 

We overlay the out-of-plane surface displacements as a function of time for three specified longitudinal positions (Fig. ~\ref{Figure3}A), and as a function of position for three specified times (Fig.~\ref{Figure3}B). In Fig.~\ref{Figure3}, the solid and dashed lines correspond to the dynamic wrinkling experiment and simulation, respectively. In both panels, it can be seen that displacement pulse broadens as it propagates, which we expect is caused by viscoelastic, dissipative, and lateral inertia effects. Wrinkles also appear to form earlier in space, closer to the striker boundary, in the simulations than in the experiments, and earlier on the crest of the propagating low-frequency pulse. 

To separate the wrinkling dynamics from that of the low-frequency pulse traveling through the substrate, we apply a spatial filter to the measured and simulated out-of-plane displacements, and subtract the filtered signal from the original, as is shown in Fig. ~\ref{Figure4}A and B, respectively. Several features become more apparent in contrast to Fig.~\ref{Figure2}. First, the wrinkles do not form near the impactor boundary in the experiments, as was shown for the quasi-static tests of Figs.~\ref{Figure1}A and B. This is also similar for experiments where the sample is flipped, such that the impactor strikes what is currently the right side (see Fig. S1), and is true, to a lesser extent, in the simulations after $\sim3$ ms. The wrinkle formation near the boundaries in the simulations also varies depending on the material properties chosen, the boundary conditions, and the parameters of the wrinkling model, including the depth used to calculate the average strain. The distance from the boundary wherein wrinkles do not form also qualitatively matches between the dynamic experiments and the quasi-static experiments for a strain of $\sim\epsilon_S$ (see Fig. S3). As such, we expect that factors such as boundary effects, including large amounts of induced curvature near the boundaries \cite{Hutchinson2011}, play a role. 

The second feature that appears by comparing Figs.~\ref{Figure4}A and B, is the wrinkle formation speed is relatively closer to the substrate sound speed in simulation than it is in the experiments (the dotted magenta line in Fig.~\ref{Figure4}A, denoting a wrinkle formation speed of $32$ m/s, is slower than the dashed magenta line denoting the substrate wave speed $c$, whereas in Fig.~\ref{Figure4}B, the wrinkle formation speed matches the dotted white line denoting speed $c_S$). We suggest that viscoelasticity may contribute to the observed delayed wrinkle formation. While viscoelasticity is taken into account for the substrate in the FE simulations, it is not incorporated into the analytical wrinkling model. Viscoelastic stiffening of the substrate material would result in an increase in the critical strain needed to induce wrinkles (by reducing the elastic mismatch between the substrate in the film), which could result in delayed wrinkle formation in regimes with strain rates approaching $1/\tau$. For the SLS model, we expect the effective material modulus to approach the instantaneous modulus ($E_0+E_1$) when the strain rates are large, and the long-term elastic modulus ($E_0$) when the strain rates are low. This would affect the critical wrinkling strain, which can be estimated as $\epsilon_c=\frac{1}{4}(3 E_s/E_f)^{2/3}$, where $E_s$ and $E_f$ are the substrate and film elastic moduli, respectively, adjusted for plane strain conditions \cite{Stafford2011}. Using the maximum and minimum expected elastic moduli, this would result in a minimum critical strain of $\epsilon_c$ and a maximum critical strain of $\epsilon_{c,i}=18 \epsilon_c$. By inspecting Fig.~\ref{Figure4}C, which shows the spatiotemporal evolution of the simulated depth-averaged normal strain rates $\dot{\epsilon_{xx}}$, we observe that regions near the wavefront, particularly positions near the left boundary, experience strain rates approaching $1/\tau$. Although the strain rates in Fig.~\ref{Figure4}C are truncated to improve visibility, small regions spatiotemporally near the striker impact exceed strain rates of $500$ s$^{-1}$. In contrast, the simulated normalized strain field $\epsilon_{xx}/\epsilon_c$ of Fig.~\ref{Figure4}D shows moderate strains in the wavefront region, and as such would not be expected to exceed the viscoelasticity-enhanced critical strain $\epsilon_{c,i}$. In the supplementary information (Fig. S4), we also use a related approach, where we estimate the instantaneous dynamic modulus (and subsequent critical strain) as the ratio of the instantaneous depth-averaged normal stress and strain, and find qualitatively good agreement concerning the delayed wrinkle formation. Other mechanisms may also play a synergistic role in the wrinkle formation delay. Prior works on dynamic buckling have suggested a time of $t_c=\lambda_c/c$ is needed for the wavefront to propagate a distance corresponding to the quasi-static critical buckling length and initiate the instability \cite{Lindberg1965,Hutchinson2006,Villermaux2007}. In our case, this results in a time of $t_c=45$ $\mu$s. However, this uniform delay would not capture the increase in the delay time as the wavefront propagates. Another potential mechanism is the inertial dynamics of the stiff surface film. Prior studies have explored the dispersion of axially compressed beams on elastic foundations \cite{Nielsen1991} and the related snapping dynamics between buckled states in axially compressed beams without elastic foundations \cite{Holmes2014}. For our case, in the linear limit, assuming an axially compressed elastic beam with a Winkler foundation \cite{Biot1937}, the frequency for a mode with wavelength equal to the critical wrinkling wavelength is zero, which suggests the dynamics of the film do not play a major role. However, the effect of nonlinear, large-deformation and viscoelastic phenomena on the inertial dynamics of films on Winkler elastic foundations present potentially rich topics for future study.  

\begin{figure}[t]
\begin{center}
\includegraphics[width=8 cm]{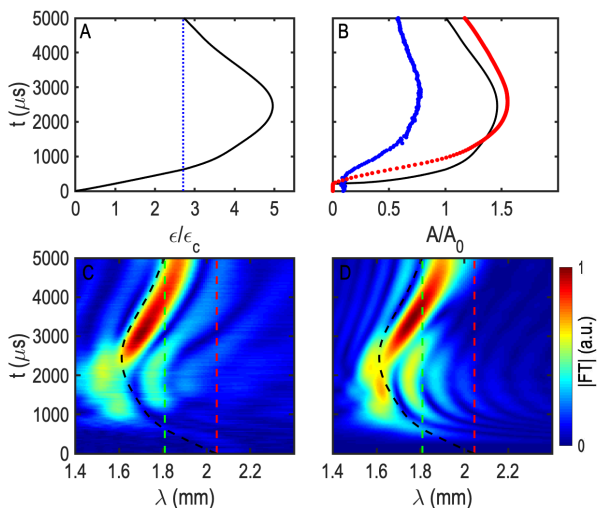}
\end{center}
\caption{\label{Figure5} [Color online] (A) Evolution of the compressive strain based on the sample boundary motion modified to account for surface curvature (see Fig. S5), normalized to $\epsilon_c$. The dotted blue line is the strain after the striker has come to rest, $\epsilon_R$. (B) Variation of the normalized wrinkling amplitude $A/A_{0}$ with time. The colors denote the wrinkle amplitude experimentally measured (blue), simulated (red), and calculated (black) using the analytical wrinkling model assuming the strain variation in panel A. Evolution of the wavelength spectra of the (C) experimentally measured and (D) simulated wrinkle pattern, obtained via Fourier transform and normalized to their respective maximum amplitude. The dashed red line is $\lambda_c$ and the dashed green line the wavelength predicted for strain $\epsilon_R$.}
\end{figure}

In the simulated strain rate field of Fig.~\ref{Figure4}C, we see small discontinuities propagating at speeds faster than the previously identified wavefront speed $c_S$, which originate from the point of striker contact with, and separation from, the sample. These discontinuities agree well with a speed of $c_L=\sqrt{(K_i+4\mu_i/3)/\rho}=424$ m/s, where $K_i=(E_0+E_1)/(3(1-2\nu))$ and $\mu_i=\mu_0+\mu_1$ are the instantaneous bulk and shear moduli, respectively. The same wavefronts can also be seen in the stress field (Fig. S4). However, as can be seen in the strain field of Fig.~\ref{Figure4}D, these wavefronts do not induce substantial longitudinal strain. 

Considering a window from $x=16$ to $49$ mm, we study the temporal variation in wrinkle amplitude and wavelength using Hilbert and Fourier transforms, respectively, and contrast it to what would be expected assuming a quasi-static compression that follows the boundary motion measured in the dynamic test. In the Fourier and Hilbert analysis of the dynamic signal, pixels where edge detection failed were replaced with zeros. Figure ~\ref{Figure5}A shows the strain $\epsilon$ expected in a quasi-static context based on the displacement of the sample boundary (see Fig. S5). The blue dotted line is the tangential normal strain induced when the striker is fully at rest against the stoppers (which we denote as $\epsilon_R$, where $\epsilon_R \simeq \epsilon_S$). Using this strain denoted by the black line as an input for the analytical wrinkling model in Eq.~\ref{wavelengthandamp}, we expect a corresponding variation in the wrinkle amplitude and wavelength, as is denoted by the black lines in Figs.~\ref{Figure5}B (solid) and C,D (dashed), respectively. 

Figure \ref{Figure5}B shows the temporal variation in wrinkle amplitude, normalized by the wrinkling amplitude, $A_0$, predicted using Eq. ~\ref{wavelengthandamp} assuming a strain of $\epsilon_R$. As can be seen by the black line in Fig. \ref{Figure5}B, wrinkling is predicted to commence when the critical strain $\epsilon_c$ has been reached, increase in amplitude with increasing compression, and then reduce as the sample separates from the striker and continues to deform. However, this simplified picture does not take into account the wave dynamics in the system. The blue and red markers in Fig. \ref{Figure5}B denote the experimentally measured and simulated wrinkling amplitude evolution obtained via Hilbert transform of the surface displacements, respectively. Their shape generally follows that of the black line, with the exception of several features. In both cases, delayed wrinkle formation can be observed in contrast to the black line, which is a result of the time required for the wavefront to propagate across the sample, as the Hilbert transform constitutes an average across the entire domain, in which wrinkles have only been partially formed. The wrinkling amplitude is also over $50\%$ smaller in experiments than is predicted from the simulations. Recalling Eq.~\ref{wavelengthandamp}, the increase in substrate modulus that is expected as a result of viscoelasticity and the high strain rates experienced as waves propagate through the sample would also translate into reduced wrinkling amplitude. The fitted wrinkling model used to calculate $A_0$ may also contribute to this disparity, as the model overpredicts the wrinkling amplitude measured in quasi-static tests by $\sim 10\%$ for strains near $\epsilon_R$. Changes in the material properties with time, for instance caused by slow curing effects, may also contribute to this disparity. To address this, we conducted a second quasi-static compression test, after the presented dynamic tests had been performed. The measured amplitudes of both tests are in close agreement, as can be seen in Fig. S6, with the exception of a region of relatively small strains (from $\epsilon \sim 0.04$ to $\epsilon \sim 0.08$), wherein the amplitude measured in the second quasi-static compression test is reduced by over $30\%$ compared to the first. However, because this difference in amplitude measured in quasi-static compression is limited to a narrow range of strains, we suggest that this effect does not account for the disparity observed in the dynamic compression tests.

Figures \ref{Figure5}C and D show the temporal evolution of the wrinkling wavelength for the experiment and simulation obtained via Fourier transform. By following the wrinkling amplitude maxima, we see that the wavelength approximately follows the trend predicted by the dashed black line. The side lobes surrounding the maxima are a result of zero padding the wrinkling signal. This agreement between simulation and experiment with respect to the wrinkling wavelength lies in contrast to the differences, which we attributed in part to viscoelastic effects, that were observed for the wrinkling amplitude and speed of wrinkling onset. Similar to the critical wrinkling strain, the critical wrinkling wavelength could also be expected to vary with changes in the substrate modulus, where $\lambda_c=2 \pi h (E_f/E_s)^{1/3}$ \cite{Stafford2011}. However, prior studies have suggested that systems experiencing instability will tend to maintain a wavelength once selected \cite{Hutchinson2011,Hutchinson2004}.  

\section{Conclusion}
In this work, we explored the response of soft elastomers containing stiff surface films that are subjected to high-speed plate impact, and observed the resulting wrinkling dynamics. The complex dynamics observed in this work suggest a need for enhanced wrinkling models that incorporate coupled inertial, viscoelastic, and large-deformation, nonlinear effects, as well as improved descriptions of how heterogeneous and non-uniaxial strain fields affect wrinkling morphology. This work encourages further studies of the dynamics of surface instabilities in soft materials, including the dynamic formation and propagation of large-deformation morphologies such as creases \cite{Hayward2008}, ridges, folds, kinks \cite{Johnson1974}, many of which are not present in related systems such as elastic films floating on liquids \cite{Witten2013}. Such dynamically evolving morphologies may find applications in areas including: impact mitigation (e.g. energy transport from impact sites in nonlinear systems) and the understanding of impact in layered, biological materials; soft electronics, particularly in the context of, potentially nonlinear, acoustic wave signal processing; and the dynamics of soft sandwich composites, such as within the context of flapping and fluttering aerospace structures. In particular, we expect that higher-amplitude, higher-dimensional, and multi-axial loading scenarios will yield rich nonlinear dynamics under high-speed impact. 

\section{Acknowledgements}We thank A. Alfaris and M. Bassford for contributions to the sample fabrication process and drop tower construction, A. Khanolkar and S. Liu for contributions to drop tower construction, and A. Emery for useful discussions. This work was supported by the the US National Science Foundation (grant no. CMMI-1536406). B.L. and L.P acknowledge support from the Washington NASA Space Grant Consortium Summer Undergraduate Research Program.

\section{Author Contributions} N.B. and S.C. designed the research. M.A.G. led the experimental effort, assisted by B.L., L.P., and M.G.. X.L. led the simulation effort, assisted by T. W.. M.A.G., B.L., and N.B. wrote the manuscript. All authors edited the manuscript.


\begin{thebibliography}{99}
\bibitem{Footnote1}
For the control experiment, the striker velocity was measured to be $V_S=10.3$ m/s and the distance the sample compressed before the impactor struck the stoppers to be $6.39$ mm. For the dynamic wrinkling experiment, the striker velocity was measured to be $V_S=9.9$ m/s and the distance the sample compressed before the impactor struck the stoppers to be $6.14$ mm.
\end{thebibliography}
\end{document}